\documentclass[prd,a4paper,showpacs,nofootinbib]{revtex4}

\usepackage{epsfig}
\usepackage{amsmath}

\newcommand{\be}{\begin{equation}}
\newcommand{\ee}{\end{equation}}
\newcommand{\ba}{\begin{eqnarray}}
\newcommand{\ea}{\end{eqnarray}}

\newcommand{\simorder}{\raisebox{-4pt}{$\, \stackrel{\textstyle >}{\sim} \,$}}
\newcommand{\simorderr}{\raisebox{-4pt}{$\, \stackrel{\textstyle <}{\sim} \,$}}

\graphicspath{{plots/}}

\begin{document}

\title{Geometric Scaling at RHIC and LHC}

\author{Dani\"el Boer}
\email{D.Boer@few.vu.nl}
\affiliation{Department of Physics and Astronomy,
Vrije Universiteit Amsterdam, \\
De Boelelaan 1081, 1081 HV Amsterdam, The Netherlands}

\author{Andre Utermann}
\email{A.Utermann@few.vu.nl}
\affiliation{Department of Physics and Astronomy,
Vrije Universiteit Amsterdam, \\
De Boelelaan 1081, 1081 HV Amsterdam, The Netherlands}

\author{Erik Wessels}
\email{E.Wessels@few.vu.nl}
\affiliation{Department of Physics and Astronomy,
Vrije Universiteit Amsterdam, \\
De Boelelaan 1081, 1081 HV Amsterdam, The Netherlands}

\date{\today}

\begin{abstract}
  We present a new phenomenological model of the dipole scattering amplitude
  to demonstrate that the RHIC data for hadron production in $d$-$Au$
  collisions for all available rapidities are compatible with geometric
  scaling, just like the small-$x$ inclusive DIS data. A detailed comparison
  with earlier geometric scaling violating models of the dipole scattering
  amplitude in terms of an anomalous dimension $\gamma$ is made. In order to
  establish whether the geometric scaling violations expected from small-$x$
  evolution equations are present in the data a much larger range in
  transverse momentum and rapidity must be probed. Predictions for hadron
  production in $p$-$Pb$ and $p$-$p$ collisions at LHC are given. We point
  out that the fall-off of the transverse momentum distribution at LHC is a
  sensitive probe of the variation of $\gamma$ in a region where $x$ is much
  smaller than at RHIC. In this way, the expectation for the rise of $\gamma$ 
  from small-$x$ evolution can be tested.
\end{abstract}


\maketitle


\section{Introduction}

The observed phenomenon of geometric scaling, i.e.\
the property that the small-$x$ DIS cross section depends only on $x$
and $Q^2$ through the combination $Q^2/Q_s^2(x)$, where $Q_s(x)$ is
referred to as the saturation scale, still requires a satisfactory
explanation. It is a property that appears in a natural way in the
(asymptotic) solutions of nonlinear evolution equations, such as the
GLR equation \cite{Gribov:1984tu,Laenen:1995fh} or the BK equation
\cite{Balitsky:1995ub,Kovchegov:1999yj}, that are expected to become
relevant at small $x$. Nevertheless, the question remains whether the
observed DIS data are obtained at sufficiently small $x$ values for
such evolution equations to be applicable. Here we want to investigate
an extension of this question to the RHIC $d$-$Au$ data and the future
LHC $p$-$Pb$ and $p$-$p$ data.  Like the DIS cross section, the hadron
production cross sections in nucleon-nucleus scattering at high
energies have been expressed in terms of the scattering of a color
dipole off small-$x$ partons, which are predominantly gluons
\cite{Mueller:1989st,Dumitru:2002qt}.  This dipole scattering
amplitude is the quantity that is expected to display geometric
scaling\footnote{More precisely, in momentum space $Q^2$ times
  the scattering amplitude is the scaling dimensionless quantity.}
and therefore should be a function of $Q^2/Q_s^2(x)$. In
nucleon-nucleus collisions the role of $Q$ is played by the transverse
momentum $q_t$ of the produced parton that fragments into the observed
hadron (or by the inverse of its Fourier conjugate $r_t$). 
For earlier works about $d$-$Au$ collisions and saturation physics we refer to
Refs.~\cite{Kharzeev:2002pc,Kharzeev:2003wz,Albacete:2003iq,Baier:2003hr,Kharzeev:2002ei}
and the review~\cite{JK}.

A successful phenomenological study of experimental DIS
data using a model for the dipole cross section was performed by
Golec-Biernat and W\"{u}sthoff (GBW)~\cite{GBW}. They found that the 
HERA data on the structure function $F_2$ at low $x$ ($x \simorderr 0.01$)
could be described well by a dipole cross section
of the form $\sigma = \sigma_0 N_{GBW}(r_t,x)$, where $\sigma_0 \simeq
23\;{\rm mb}$ and the scattering
amplitude $N_{GBW}$ is given by 
\be
N_{GBW}({r}_t,x) = 1-\exp\left(-\frac{1}{4} r_t^2
Q_s^2(x) \right). 
\label{NGBW}
\ee
This amplitude depends on $x$ and $r_t$ (the transverse size of the
dipole) only through the combination $r_t^2Q_s^2(x)$, which means it is 
geometrically scaling.  
The $x$-dependence of the saturation scale is given by
\be
Q_s(x) = 1\,\mathrm{GeV}\,
\left(\frac{x_0}{x}\right)^{\lambda/2}, 
\label{Qsx2}
\ee
with $x_0 \simeq 3 \times 10^{-4}$ and $\lambda \simeq 0.3$. For nuclear
targets $Q_s^2$ contains an additional factor $A^{1/3}$.

Model independent analyses of the HERA data show that the low-$x$ data display
geometric scaling for all $Q^2$~\cite{Stasto:2000er,Gelis:2006bs}, even though
the GBW model, (\ref{NGBW})-(\ref{Qsx2}), was found to be
inconsistent with newer, more accurate data
at large $Q^2$ and requires modification. In Ref.\ \cite{Bartels:2002cj} such 
a modification was proposed which includes DGLAP evolution, in order 
to fit the $Q^2 > 20$ GeV$^2$ data. 
In Ref.\ \cite{Kwiecinski:2002ep} the impact of DGLAP 
evolution on a geometric scaling solution has been numerically studied. 
An initial
condition was constructed so that at $Q^2=Q_s^2(x)$, with $Q_s(x)$ as in 
Eq.\ (\ref{Qsx2}), 
the dipole cross section at leading order (and hence
$\alpha_s(Q^2) x g(x,Q^2)/Q^2$) is a constant as required for a geometric
scaling solution. 
It was found that under DGLAP evolution to higher values of
$Q^2$ geometric
scaling is not violated for $\lambda \geq 4 N_c \alpha_s/\pi$ in the fixed
coupling constant case and only mildly violated for all values of $\lambda$ 
in the running coupling constant case. In the latter case geometric scaling (GS) 
holds to very good approximation in the region 
$\log Q^2/Q_s^2 \ll \log Q_s^2/\Lambda^2$  
(here $\Lambda$ denotes $\Lambda_{\rm QCD}$). 
One can conclude that although the DGLAP evolution equation for
the gluon distribution does not necessarily lead to a GS solution itself, 
this GS property can be preserved to a large extent. The DGLAP induced 
violations of GS can remain small over a wide range of $Q^2$ values. 

Similar studies have been performed for the BFKL equation
\cite{Kuraev:1977fs,Balitsky:1978ic}.  The solution of the BFKL equation
with an appropriate boundary condition at $Q_s$ was found to be
geometrically scaling in leading order in the saddle point
approximation
\cite{LevinTuchin,MuellerTr,Triantafyllopoulos:2002nz,IIM2}.  
 Beyond leading order for $1 \simorderr \log Q^2/Q_s^2 \ll
\log Q_s^2/\Lambda^2$ this solution shows approximate scaling. At the
scale $Q_{gs} \equiv Q_s^2/\Lambda$ the violations of geometric
scaling are considered sizeable.  This of course assumes that the BFKL
equation governs the evolution in this entire region that has been
called the extended geometric scaling (EGS) region. 
The region $Q^2<Q_s^2$ is referred to as the saturation region.

The EGS region need not be equal to the region in which GS is
  observed in experiments, since it is unclear that the chosen
  evolution equation is appropriate in the entire region in the first
  place.  But even if this is the case, one could not
determine $Q_{gs}$ from the data for a given rapidity.  Only if one
studies the data as function of $Q^2/Q_s^2$ for a range of rapidities
(for which $Q_{gs}$ is not a constant scale) will one be able to
establish the extent to which GS is violated.
 
As mentioned, the DIS data for $Q^2 > 20$ GeV$^2$ prompted the authors of
Ref.\ \cite{Bartels:2002cj} to propose a modification of the GBW model
which includes DGLAP evolution.  However, in Ref.~\cite{IIM} a model
(IIM) has been put forward that is a modification of the GBW model and
incorporates the violations of geometric scaling expected to arise
from BFKL evolution in the EGS region. This model leads to a satisfactory fit 
to DIS data, but without the need to include DGLAP evolution at larger $Q^2$.  
In Ref.\ \cite{Gotsman:2002yy} a description of the DIS data is obtained by 
taking into account both BK and DGLAP evolution. The fact that DIS data can be described using 
different approaches suggests that the small-$x$ DIS data do not span a sufficiently large
region in $Q^2$ and $x$ to discriminate between the different types of evolution. 
The question is whether or not the RHIC and future LHC data do span 
a sufficiently large region.

In order to investigate GS violations in the RHIC data, in Refs.\ \cite{DHJ1,DHJ2} a 
phenomenological model, similar to the IIM model, has been put forward 
(following in part the earlier study of Ref.\ \cite{KKT} based on \cite{Kharzeev:2003wz}). 
We will refer to this model as the DHJ model. It offers a good description 
of the $p_t$ distribution of hadrons produced in $d$-$Au$ collisions  at RHIC in 
the forward region \footnote{As it turned out the central-rapidity study of 
Ref.\ \cite{DHJ2} contained an error in the numerical code. The larger 
$p_t$ data for $y_h=0, 1$ are in fact not well-described by 
the DHJ model as will be seen.}, and even in $p$-$p$ collisions in the
very forward rapidity region \cite{BDH}. 
 
According to Refs.\ \cite{DHJ1,DHJ2}  the cross section\footnote{To be 
more precise, Eq.\ (\ref{eq:conv2}) is an expression for the minimum bias
  invariant yield.} of single-inclusive forward hadron production 
in high-energy nucleon-nucleus collisions is described in terms of the 
dipole scattering amplitude in the following way,
\ba
{dN_h \over dy_h d^2p_t} &=& 
{K(y_h) \over (2\pi)^2} \int_{x_F}^{1} dx_1 \, {x_1\over x_F}
\Bigg[f_{q/p}(x_1,p_t^2)\, N_F \left({x_1\over x_F}p_t,x_2\right)\,
D_{h/q}\left({x_F\over x_1},p_t^2\right)
\nonumber \\
& &+~
f_{g/p}(x_1,p_t^2)\, N_A \left({x_1\over x_F}p_t,x_2\right)\, 
D_{h/g}\left({x_F\over x_1},p_t^2\right)\Bigg]~.
\label{eq:conv2}
\ea 
A summation over quark flavors $q$ is understood. Here $N_F$ describes 
a quark scattering off the nucleus, while $N_A$ applies to 
a gluon. The parton distribution functions $f_{q/p}$ and the
fragmentation functions $D_{h/q}$ are considered at the scale $Q^2=p_t^2$,
which we will always take to be larger than 1 GeV$^2$. 
The momentum fraction of the target partons equals
$x_2=x_1\exp(-2y_h)$. We find that for pion, kaon, proton
and even $\Lambda$ production we can to good approximation neglect
finite mass effects, i.e.\ we equate the
pseudorapidity $\eta$ and the rapidity $y_h$ and use
$x_F=\sqrt{p_t^2+m^2}/\sqrt{s}\exp(\eta)\approx
p_t/\sqrt{s}\exp(y_h)$.
Finally, there is an overall $K$-factor that effectively accounts
  for NLO corrections. As it is expected that these corrections are
  more important at small $y_h$, the $K$-factor is allowed to be $y_h$
  dependent.  A NLO pQCD
  analysis of the process $p \, p \to \pi^0 \, X$ at mid-rapidity for
  RHIC energies shows that such $K$-factors are relatively constant
  with $p_t$ \cite{Jager:2002xm}.

The dipole scattering amplitude of the DHJ model is
given by~\cite{DHJ1,DHJ2}:
\ba 
N_A({q}_t,x_2) & \equiv & \int d^2 r_t\: e^{i \vec{q}_t
    \cdot \vec{r}_t} N_A(r_t,q_t,x_2)\nonumber\\
& \equiv & \int d^2 r_t\: e^{i \vec{q}_t
    \cdot \vec{r}_t} \left[1-\exp\left(-\frac{1}{4}(r_t^2
      Q_s^2(x_2))^{\gamma(q_t,x_2)}\right) \right]~.
\label{NA_param}
\ea
Note that $\gamma$ is a function of $q_t$ rather than $r_t$. This allows one to
compute the Fourier transform more easily. 
The corresponding expression $N_F$ for quarks is obtained from $N_A$
by the replacement $(r_t^2Q_s^2)^\gamma \to ((C_F/C_A) r_t^2Q_s^2)^\gamma$,  
with $C_F/C_A=4/9$.
The exponent $\gamma$ is usually referred to as the
``anomalous dimension'', although the connection between $N_{A/F}$ and
the gluon distribution inside the nucleus cannot always be made.

The anomalous dimension of the DHJ model is parameterized as
\be
\gamma(q_t,x_2) =
\gamma_s + (1-\gamma_s)\, \frac{\log(q_t^2/Q_s^2(x_2))}{\lambda
y+d\sqrt{y}+\log(q_t^2/Q_s^2(x_2))}, \label{gammaparam}
\ee
where $y=\log 1/x_2$ is minus the rapidity of the target parton. The
saturation scale $Q_s(x_2)$ and the parameter $\lambda$ are taken from the GBW
model, as given in Eq.\ (\ref{Qsx2}). Here $Q_s$ includes the additional 
factor $A^{1/3}$, for which DHJ use $A_{\rm eff} =18.5$ in the $d$-$Au$ case. 
The parameter $d$ was fitted to the data and set to $d=1.2$.
This choice of $\gamma$ leads to a
geometric scaling solution at $q_t=Q_s$ where $\gamma=\gamma_s=0.628$ and
incorporates to a certain extent the violation expected from BFKL evolution
for larger $q_t$. The anomalous dimension of DHJ is of the form
$\gamma=\gamma_s+\Delta\gamma$, where the scaling violations arising from
$\Delta\gamma$ behave as $\log(q_t^2/Q_s^2)/y$ for large $y$ and $q_t^2
\simorder Q_s^2$ as resulting from the analyses of Refs.\
\cite{MuellerTr,Triantafyllopoulos:2002nz,IIM2}. The question we will address
in this paper is whether these violations are really seen in the available
data. The fact that the DHJ model works well for forward hadron production in
$d$-$Au$ collisions does not demonstrate that there are actually violations
present as we will show in detail.

For very large $q_t$, or equivalently for small $r_t$, 
one can use to good approximation 
\be 
N_A({r}_t,q_t,x_2) \approx
\frac{1}{4}(r_t^2 Q_s^2(x_2))^{\gamma(q_t,x_2)}.
\label{NgammaApprox}
\ee
DHJ used the perturbative $t$-channel one-gluon exchange result  
to conclude that $\gamma\to1$ as $q_t \to \infty$. 
However, as discussed in Ref.\ \cite{BUW} also if the BFKL equation
governs the large-$q_t$ region, one can find that $\gamma \to 1$ at large $q_t$. 
The way in which  $\gamma$ approaches 1 directly  
determines how fast the cross section will fall off with increasing $p_t$ as
we will discuss in the next section. 

In the DHJ model one retains GS approximately when $\Delta \gamma$ is small w.r.t.\ $\gamma_s$. 
For large, but fixed rapidity ($y \gg (d/\lambda)^2$) this holds numerically up to 
$q_t \approx Q_{gs} = Q_s^2/\Lambda$, when $\log Q_{gs}^2/Q_s^2 \sim \lambda y$.  
In general it is not simply the variation of $\gamma$ 
that determines the GS violations. If $\gamma$ is chosen
to be a function of $q_t^2/Q_s^2$ or $r_t^2 Q_s^2$ only, the model is never GS
violating no matter how fast it approaches $1$ at large $q_t$. We will present
such a scaling model below and demonstrate that it can describe the RHIC data in both 
the central and forward rapidity regions. Although the new parameterization of
$\gamma$ is similar in form to that of the DHJ model, it does not have the 
GS violating behavior nor the logarithmic rise expected from the BFKL 
(and more generally, BK) equation.

The outline of this paper is as follows. In Section II we discuss the
properties of a new phenomenological model of the dipole cross section and
the parameters fitted to RHIC data. A comparison to DIS data and the GBW model 
is also made. 
In Section III we present predictions for hadron and jet production at LHC. 
In Section IV we summarize our main conclusions. 

\section{New model}

\label{pheno_sec} We will now use RHIC data and the cross section expression in Eq.\ 
(\ref{eq:conv2}), as employed by DHJ \cite{DHJ1,DHJ2}, to constrain the
anomalous dimension $\gamma$ entering the dipole cross section
(\ref{NA_param}). Especially, we would like to address the question
whether the RHIC data really require violation of geometric scaling
as it was claimed in Refs.\ \cite{DHJ1,DHJ2} and also stated in Ref.\ 
\cite{Iancu:2006uc}. In DIS a scaling behavior of
the dipole scattering amplitude $N(r Q_s)$ maps directly into a
scaling of the DIS cross section
$\sigma_{\gamma^\ast\,p}$, which is clearly observable in the data at small $x$ 
($x \simorderr 0.01$).
Due to the convolution in Eq.\ (\ref{eq:conv2}) the situation is more
involved in hadron-hadron collisions where no such scaling in terms of
the observed kinematic variables ($y_h$ and $p_t^{}$) can be expected. 
Therefore we have to focus
on the question whether hadron production in $d$-$Au$ collisions is
describable in terms of scaling dipole cross sections $N_{A}$ and $N_F$.

An anomalous dimension $\gamma$ leading to a geometric scaling dipole
cross section should depend on $w=q_t/Q_s(x_2)$, but not separately 
on $q_t$ and rapidity $y=\log 1/x_2$. We will take a value
$\gamma_1 = \gamma(w=1)$ of the order of $\gamma_s$ 
and $\gamma \to 1$ for larger $w$. The
parameterization that we adopted reads
\begin{equation}
 \gamma(w)=\gamma_1+(1-\gamma_1)\frac{(w^a-1)}{(w^a-1)+b}\,.
\label{gamma_alpha}
\end{equation}
The two free parameters $a$ and $b$ will be fitted to the RHIC data. 
The parameterizations (\ref{gammaparam}) and (\ref{gamma_alpha}) differ
not only in the scaling behavior, but also the large $q_t$ limit of $\gamma
\to 1$ is approached much faster in the latter case. 
This will lead to different large momentum slopes of the
dipole scattering amplitude (\ref{NA_param}) and therefore to different
predictions for the large $p_t$ slope using Eq.\ (\ref{eq:conv2}). For larger
$w=q_t/Q_s$ the exponent can be expanded and the dipole scattering
amplitude (\ref{NA_param}) simplifies, cf.\ Eq.\ (\ref{NgammaApprox}) 
(we will suppress the possible $y$ dependence),
\begin{align}
\begin{split}
  N_A({q}_t) &\approx
  \frac{2\pi}{q^2_t}\frac{1}{w^{2\gamma(w)}}\frac{1}{4}\int_0^\infty dz\, z\,{\rm
    J}_0(z)\,(-z^{2 \gamma(w)})
=\frac{2\pi\,2^{2\gamma(w)-1}\,Q_s^{2\gamma(w)}}{q_t^{2\gamma(w)+2}}\,
\frac{\Gamma(1+\gamma(w))}{-\Gamma(-\gamma(w))}
\\&
\stackrel{\gamma(w)\to 1}{\approx}
\frac{4\pi\,Q_s^2}{q_t^4}\,(1-\gamma(w))
\propto
\left\{\begin{array}{cl}
    \frac{Q_s^2}{q_t^4\log(q_t^2/Q_s^2)}& \mbox{} \quad 
\text{for $\gamma$ of Eq.\ (\ref{gammaparam})}
\\[2ex]
\frac{Q_s^{2+a}}{q_t^{4+a}}& \mbox{} \quad \text{for $\gamma$
      of Eq.\ (\ref{gamma_alpha})} 
\end{array}\right.\,.
\label{NA_asympt}
\end{split}
\end{align}
For a constant $\gamma < 1$ the amplitude will drop even more slowly than both
these models, namely $\propto Q_s^{2\gamma}/q_t^{2\gamma+2}$. For $\gamma=1$ 
(the GBW model) one
finds on the other hand an unrealistic exponential fall-off 
$\propto \exp(-q_t^2/Q_s^2)/Q_s^2$, which could be corrected by including 
a GS violating logarithm as in the MV model \cite{MV}.
Due to the convolution in Eq.\ (\ref{eq:conv2}) with the parton distribution
and fragmentation functions, the slope of the $p_t$ distribution is
not so simple to estimate. Empirically we find that the power of the
$p_t$ distribution is roughly a factor of one to two larger than the power of
the dipole scattering amplitude. 
Below we are going to determine this power. We emphasize that the fall-off with $p_t$ is not determined by
the size of the scaling violations. In order to observe such
violations one has to study both the $y_h$ and $p_t$ dependence over a
significantly large range. Moreover, the scaling properties of the dipole scattering amplitude 
are not directly visible in the hadron production data, 
due to the parton distribution and fragmentation functions. 
\subsection{Comparison with RHIC data}

\begin{figure}[htb]
\centering
\includegraphics*[width=110mm]{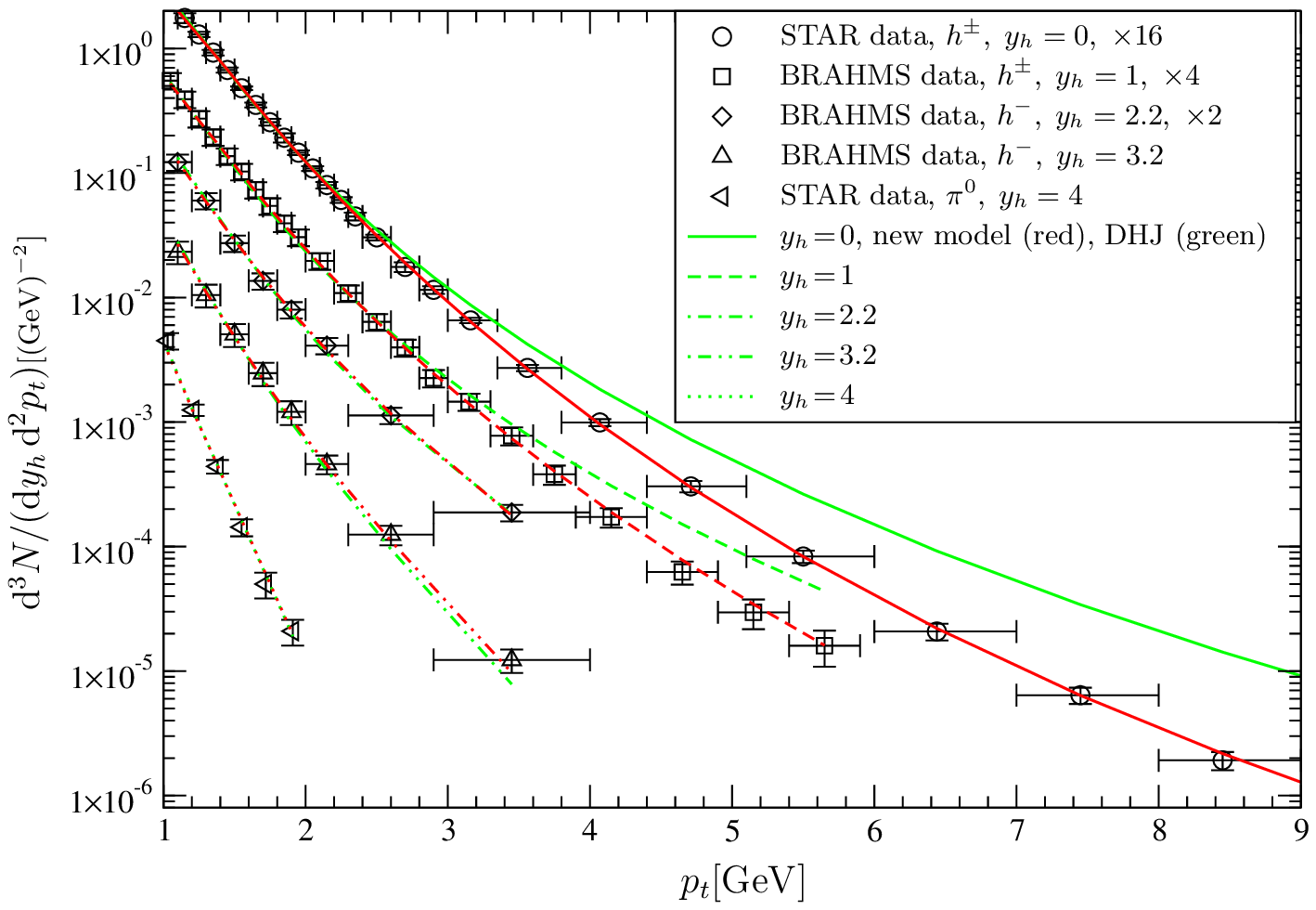}
\caption{\small\label{fig_RHIC} Transverse momentum distribution of
  produced hadrons in $d$-$Au$ collisions as measured at RHIC (black
  symbols) for various rapidities $y_h$. Using the scaling
  parameterization (\ref{gamma_alpha}) the data are well described by
  the expression (\ref{eq:conv2}) with an appropriate $K$-factor
  (red/dark curves). The DHJ model
  (\ref{gammaparam}) works well only for smaller $p_t$ (green/light
  curves).  To make the plot clearer, the data and the curves for
  $y_h=0, 1$ and $2.2$ are multiplied with arbitrary factors, namely
  16, 4 and 2, respectively.  The STAR data at $y_h=0$ are from Ref.\ \cite{Adams:2003im} 
and $y_h=4$ from \cite{Adams:2006uz}. The BRAHMS results for $y_h=1-3.2$ can be
  found in \cite{Arsene:2004ux}.}
\end{figure}

In Fig.~\ref{fig_RHIC} we show our estimate for $dN_h/(dy_h
  d^2p_t)$ that follows from the integral in Eq.\ (\ref{eq:conv2})
with our parameterization for $\gamma(w)$ (\ref{gamma_alpha}), which
enters the dipole scattering amplitude (\ref{NA_param}). All $p_t$
distributions of produced hadrons measured at RHIC in $d$-$Au$ collisions
\cite{Adams:2003im,Arsene:2004ux,Adams:2006uz} are well described. At
the saturation scale we have chosen here for $\gamma$ the same value
$\gamma_s=0.628$ as in the DHJ model. We also take $A_{\rm eff}=18.5$. 
We obtain the best fit of the data for:
\begin{equation} 
a=2.82 \quad \text{and} \quad b=168\,.
\label{bestfitparams}
\end{equation}
As mentioned, this LO analysis requires the inclusion of a $K$-factor
to account for NLO corrections, which are expected to become more
relevant towards central rapidity. Following DHJ, the $K$-factor is
allowed to vary with $y_h$, but is demanded to be $p_t$ independent.
The $K$ factors we obtain for $y_h=0,1,2.2,3.2,4$ are for our new
model equal to $K=3.4, 2.9, 2.0,1.6, 0.7$ and for the DHJ model
$K=4.3, 3.3, 2.3, 1.7, 0.7$. We have assumed isospin invariance to
obtain the parton distributions for a deuteron from those for a
proton, using the CTEQ5-LO ones \cite{Lai:1999wy}. Furthermore, we use the
KKP fragmentation functions of Ref.~\cite{Kniehl:2000fe}.

\begin{figure}[htb]
\centering
\includegraphics*[width=110mm]{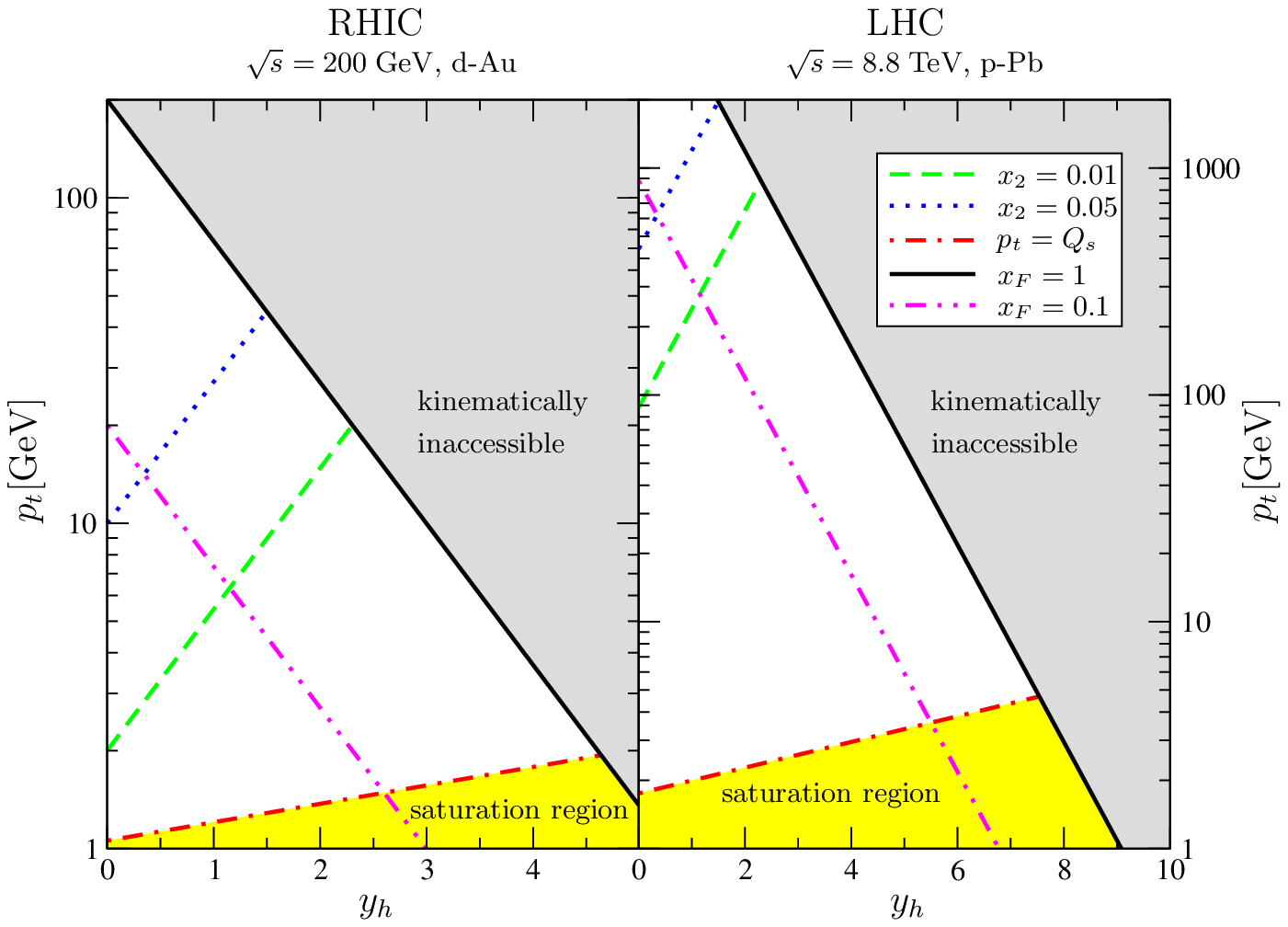}
\caption{\small\label{fig_regions} Illustration of the kinematical
  ranges relevant for RHIC and LHC. The saturation region is set by the line
  $q_t=x_1/x_F\,p_t=Q_s(x_2)$. Since the dominant contribution  to (\ref{eq:conv2})  
comes from the region $x_1$ close to $x_F=p_t/\sqrt{s}\exp(y_h)$, we used for this plot
  $x_2=x_1\,\exp(-2y_h)\approx p_t/\sqrt{s}\exp(-y_h)$ and $x_1\approx x_F$. For $d$-$Au$ 
  we have taken $A_{\rm eff}=18.5$ and for $p$-$Pb$ 20.  The curves of
  constant $x_2$ indicate the regions where small-$x$ physics may become 
relevant. }
\end{figure}

\mbox{From} this analysis we can conclude that a GS dipole scattering
amplitude is completely compatible with the data and therefore the
conclusion that GS violations are observed at RHIC cannot be drawn.
Of course, a scaling violating amplitude, i.e. a $\gamma$ that depends
on $w$ and the rapidity $y$ explicitly, is not ruled out by the data
either.  What can be concluded further is that the logarithmic rise of
$\gamma$ resulting from the BFKL evolution incorporated in the DHJ
model is ruled out in the central region, see Fig.~\ref{fig_RHIC}.
This may simply indicate that $x_2$ is already so large that one is in
the DGLAP region.  In Fig.\ \ref{fig_regions}, the kinematic region
where $x_2$ is small is indicated in terms of the observables $p_t$
and $y_h$.  Where the DHJ model starts to deviate from the data $x_2$
becomes larger than 0.01, although $Q_s$ is still larger than in DIS
at $x=0.01$. If one were to exclude the central rapidity RHIC data in
the model fit, one could also obtain a scaling model with a
logarithmically rising, or even constant, $\gamma$.

\begin{figure}[htb]
\centering
\includegraphics*[width=110mm]{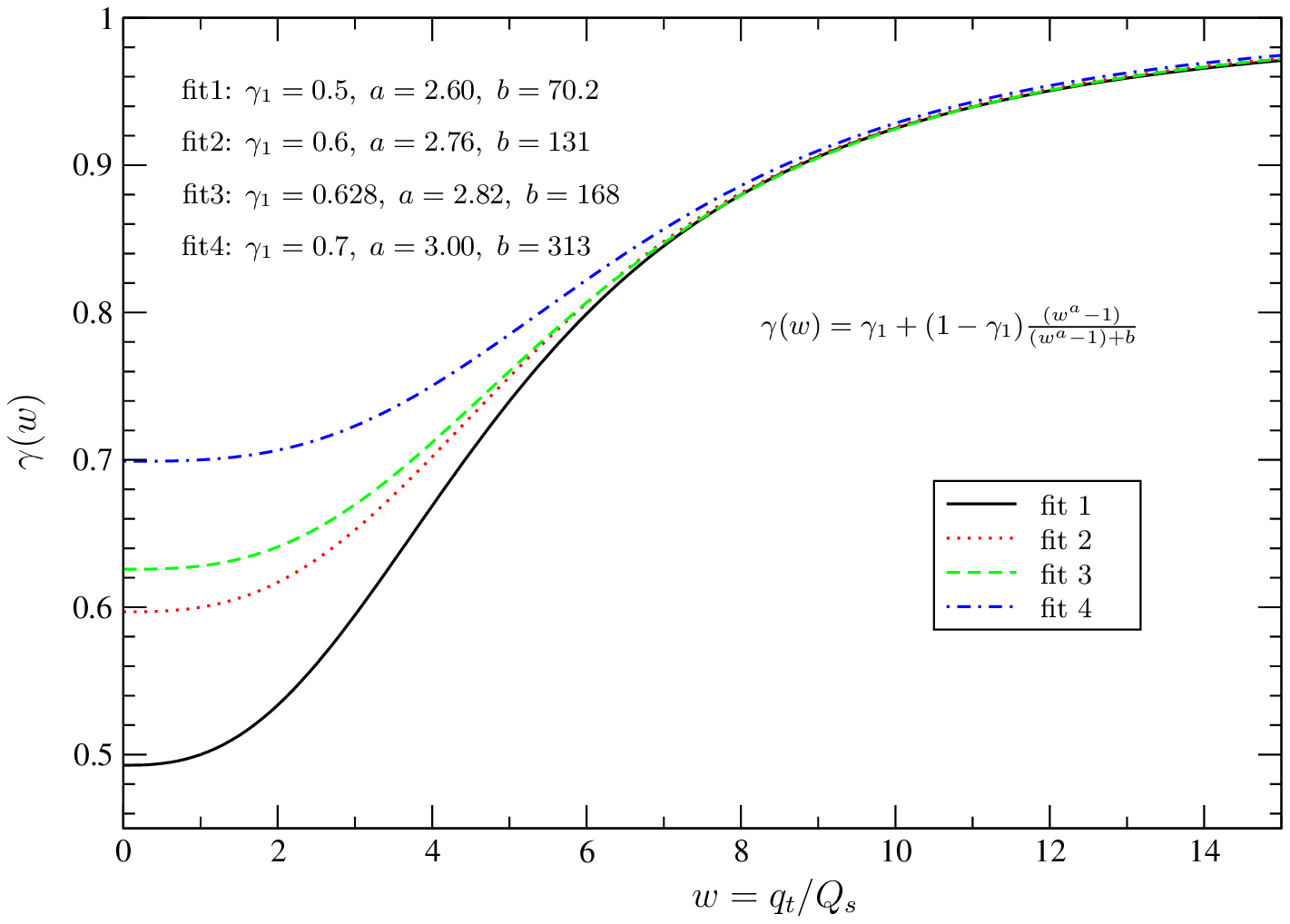}
\caption{\small\label{fig_gamma} Various fits of $\gamma(w)$, which
  describe the RHIC data equally well. For choices of $\gamma$ outside this 
range the data are less well described.}
\end{figure}

To indicate how much $\gamma$ is constrained by the RHIC data,
Fig.~\ref{fig_gamma} shows various $\gamma(w)$'s that describe the
available data equally well. They are all parameterized as in Eq.\
(\ref{gamma_alpha}) with different $a$ and $b$ values, but require
different $K$ factors.  Clearly, $\gamma$ is less well determined
close to the saturation scale than in the dilute region. This is
because the integrand entering the dipole scattering amplitude
(\ref{NA_param}) around the saturation scale $r=1/Q_s$ is only weakly
dependent on $\gamma$. In addition, the forward data ($y_h=3.2$
  and 4) are essentially sensitive only to $\gamma_1$, since they 
  probe the region where $w$ is close to 1.  Therefore, the rise of
  $\gamma$ with $w$ is effectively constrained only by the data for
  $y_h=0,1$.

It is important to realize that given a non-scaling $\gamma(w,y)$
  that fits the data for some value of $y_h$ one can always find a
  scaling $\tilde{\gamma}(w)$ that leads to the same $p_t$
  distribution.  This may not be obvious since even if $y_h$ is fixed,
  a range of $y$ values is probed in the convolution integral
  (\ref{eq:conv2}).  However, the scaling parameter $w$ can always be
  expressed as a function of $y$ and $y_h$,
 \begin{equation}
  w=\frac{x_1}{x_F}\frac{p_t}{Q_s}=x_2\exp[y_h]\frac{\sqrt{s}}{Q_s}=\exp[-y+y_h]\frac{\sqrt{s}}{Q_s(y)}=w(y;y_h,s).
\end{equation}
Hence, if $y_h$ is kept fixed one can express the rapidity $y$ in terms of
$w$ and define a scaling
$\tilde{\gamma}(w)\equiv\gamma(w,y=y(w))$ that leads to the same
results as $\gamma(w,y)$. Clearly, without probing a sufficiently
large range of $y_h$ scaling violations cannot be established. 

As mentioned before, $\gamma$ is chosen to be a function of $q_t$ rather than 
$r_t$. For completeness it should be said that it is possible to 
describe the data equally well with a scaling $\gamma$ that depends on $r_tQ_s$.
In general, this will be a different function than one would obtain by simply replacing 
$q_t$ with $1/r_t$ in $\gamma$  which would lead
to unphysical oscillations in the dipole scattering amplitude and hence in the
hadron production cross section.

\subsection{Compatibility with deep-inelastic scattering}
\label{sec_dis}
Since the parameterization (\ref{NA_param}) of the
dipole scattering amplitude uses an 
anomalous dimension $\gamma \neq 1$, 
the resulting amplitude is quite
different from the GBW model. Therefore, it is important to check
whether our anomalous dimension $\gamma(w)$ is still compatible with
the DIS data. For this we use the following expression for the dipole 
scattering amplitude
\begin{equation} 
N_\gamma({r}_t,Q,x) = 1-\exp\left(-\frac{1}{4} (r_t^2
Q_s^2(x))^{\gamma(w=\sqrt{Q^2/Q_s^2(x)})}\right)\,,
\label{Ngamma}
\end{equation}
where $Q_s$ is given by Eq.\ (\ref{Qsx2}) and for $\gamma$ we use
  our model, Eqs.\  (\ref{gamma_alpha}) and (\ref{bestfitparams}).

Following the procedure in \cite{GBW}, we predict the total cross
section $\sigma_{\gamma^\ast p}=\sigma_T+\sigma_L$  by
folding the resulting dipole cross section $\sigma=\sigma_0\,N_\gamma$
with the perturbatively calculable photon wave function,
\begin{equation}
 \sigma_{T,L}(x,Q^2)=\int dz \int d^2r_t\;\left|\psi_{T,L}(z,r_t,Q^2)\right|^2\,\sigma(r_t,x)\,,
\end{equation}
where $z$ is the longitudinal momentum fraction of the quark in the dipole.

\begin{figure}[htb]
\centering
\includegraphics*[width=110mm]{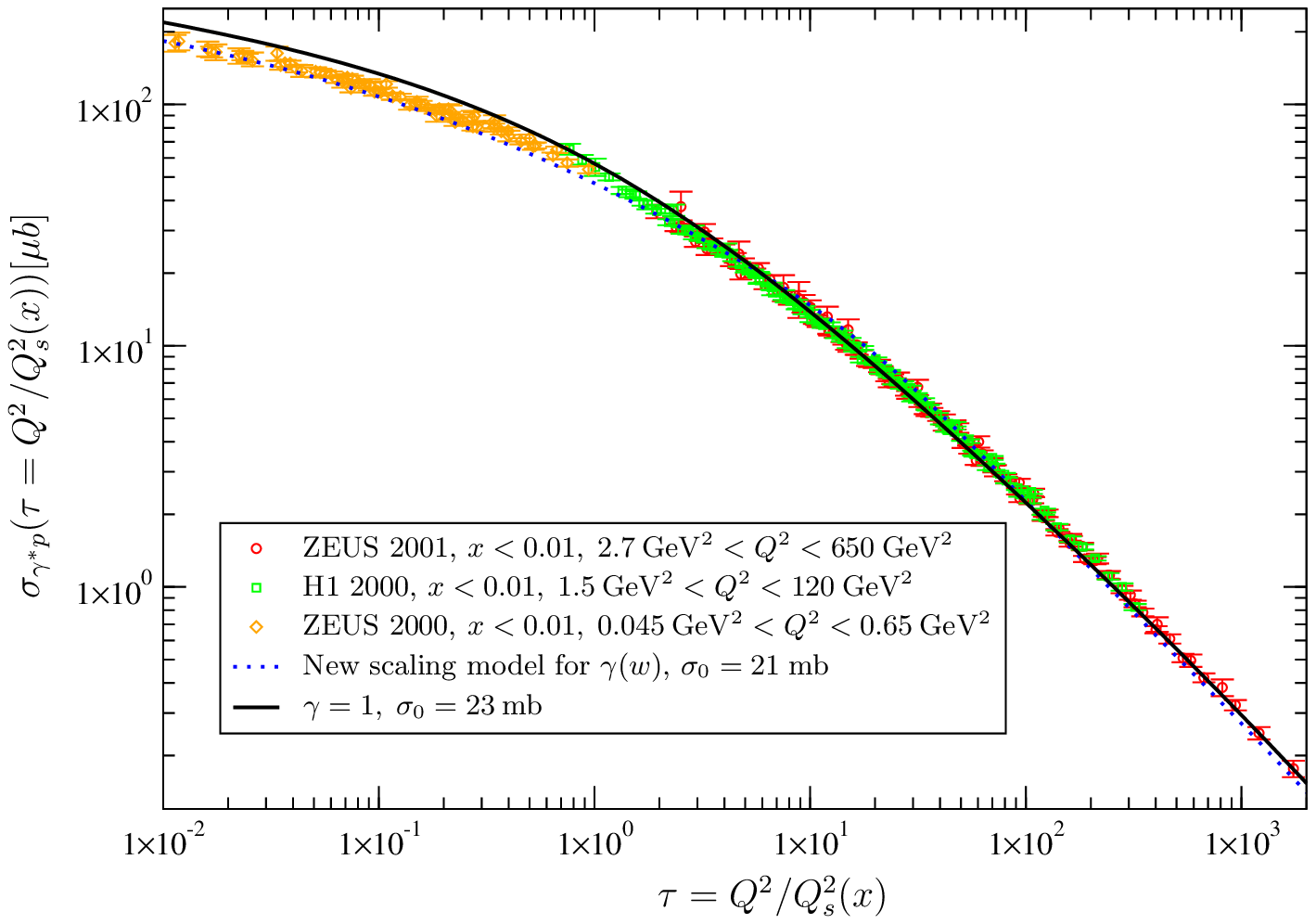}
\caption{\small\label{fig_dis} The $\gamma^\ast p$ cross section as a
   function of the scaling variable $\tau$ measured at HERA
  \cite{Adloff:2000qk,Breitweg:2000yn,Chekanov:2001qu}. We compare the
data with the predictions following from the original GBW model, where
$\gamma=1$, and the modified version fitted to RHIC  data, where $\gamma$ depends on
 $Q^2/Q_s^2(x)$. }
\end{figure}

In Fig.~\ref{fig_dis} we show the small-$x$ HERA data
\cite{Adloff:2000qk,Breitweg:2000yn,Chekanov:2001qu} in a large
kinematic range as a function of $\tau=Q^2/Q_s^2(x)$. Following Ref.\
\cite{Bartels:2002cj}, we scale the H1 data \cite{Adloff:2000qk} by a
factor 1.05, which is consistent with the normalization uncertainty.
As can be seen, the data for $x<0.01$ depend on $x$ and $Q^2$ only
through the variable $\tau$. In Fig.~\ref{fig_dis} we compare these
data with the original GBW model and the prediction following from our
modified $\gamma$ obtained from a fit to RHIC data.  For both models
we have neglected effects from finite quark masses in the photon wave
function, which break geometric scaling. As a result the cross section
of the GBW model overshoots the data at small $\tau$, i.e.\ at small
$Q^2$. Using the modified $\gamma$ this effect and therefore the
fitted quark mass is smaller since the smaller value of $\gamma$ in
the saturation region suppresses the cross section.  Further details
of the small-$\tau$ behavior can be found in e.g.\ \cite{Avsar:2007ht}.
In addition, we use a somewhat smaller $\sigma_0$ value in order to
obtain a better description of the data.  No parameters of $\gamma$
are tuned.  The smaller value of $\sigma_0$ is forced by the region
$\tau\approx 10\ldots 100$ where $\gamma$ is not yet close to one but
the effective value of $r_tQ_s$ is already large. Given the
normalization uncertainty of our model, we do not consider the smaller
value of $\sigma_0$ a problem. Of course, it would be possible to
obtain an optimized parameterization of $\gamma$ by a simultaneous fit
to the RHIC and DIS data.  But at this stage we conclude that the
model, which we constructed to describe the RHIC data, can describe
the DIS data equally well as the GBW model if one adjusts the
additional parameter $\sigma_0$.

  We end this section with a comment on whether or not the factor
  $(C_F/C_A)$ should have been included in $N_\gamma$ of Eq.\ (\ref{Ngamma}),
  as was done for $N_F$ earlier. In order to compare the DHJ model or our new 
  model with the GBW model, it would indeed be better to use Eq.\ 
  (\ref{NA_param}) as a model for $N_F$ and scale $(r_t^2Q_s^2)^\gamma 
  \to ((C_A/C_F) r_t^2Q_s^2)^\gamma$ to obtain $N_A$. A fit to RHIC data would
  then result in a somewhat different $\gamma$. Since this is not done by DHJ
  and we are specifically interested in a comparison to the DHJ model, 
  we follow DHJ's approach. This does however obscure the comparison to the
  GBW model somewhat, since that is a model for $N_F$ without the factor
  $C_F/C_A$. Note that for models with $\gamma \neq 1$ this cannot be accounted for 
by rescaling $Q_0$, because one does not scale $Q_s$ that enters in 
  $\gamma$. In a future combined fit to RHIC and DIS data one would of course 
  like to avoid this slight conceptual discrepancy.

\section{LHC predictions}

\subsection{Hadron production}

We have seen that where
  the DHJ model curves deviate from the RHIC data, the $x_2$-values
  probed are not very small. However, at LHC due to the much higher
  energies, the region of small $x_2$ extends to a much larger range
  of $p_t$, so that the predictions of the DHJ model and the new model
  will be different even at small $x_2$.  In Fig.\ \ref{fig_regions}
  the region of small $x_2$ is depicted in terms of $p_t$ and $y_h$
  for $p$-$Pb$ collisions at LHC.  In this section we discuss the
  predictions following from the DHJ model and our new scaling model for the hadron
  production cross section for $p$-$p$ collisions at $\sqrt{s}=14$ TeV
  and for $p$-$Pb$ collisions at $\sqrt{s}=8.8\;{\rm TeV}$.
 
\begin{figure}[htb]
\centering
\includegraphics*[width=110mm]{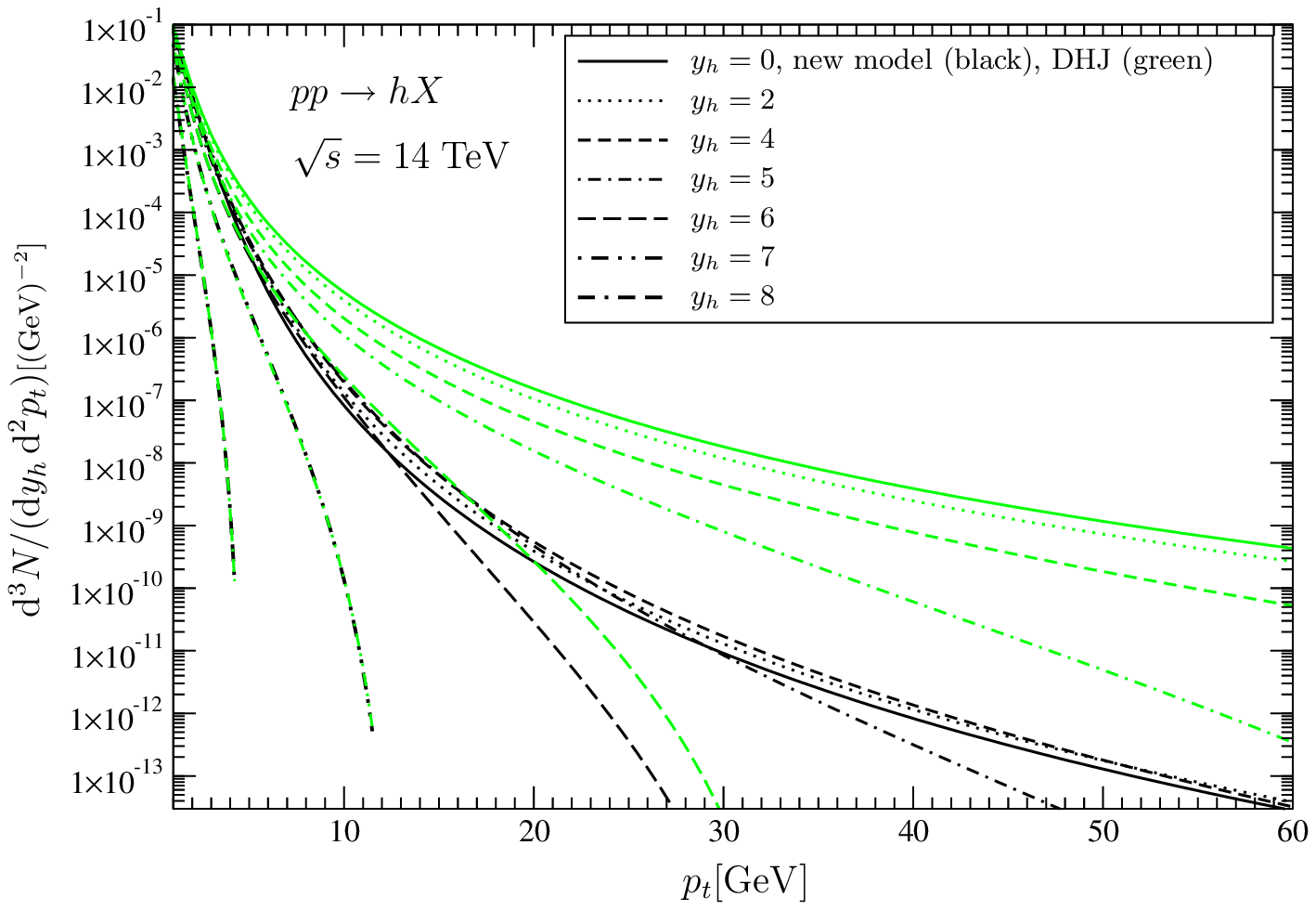}
\caption{\small\label{fig_LHC} Predictions of the transverse momentum
  distributions of produced hadrons in $p$-$p$ collisions at the LHC energy of
  $\sqrt{s}=14\;{\rm TeV}$ and various rapidities $y_h=0-8$. The
  distributions from the scaling model are represented by the black lines and those
from the DHJ model by the green/light ones.}
\end{figure}

  In Fig.~\ref{fig_LHC} we show the predictions for the $p$-$p$
  collisions at $\sqrt{s}=14\;{\rm TeV}$.  For smaller $p_t$ the
  predictions of the DHJ model and the new model are comparable.  This
  can be expected since this region corresponds to the small-$p_t$
  region at RHIC. For larger rapidities, i.e.\ $y_h \approx 7 - 8$,
  the predictions are indistinguishable since the reachable momenta
  $q_t\le \sqrt{s}\exp(-y_h)$ are so small that the ratios $w=q_t/Q_s$
  are always so close to one that $\gamma$ is effectively equal to
  $\gamma_s$.  However, there is quite a large range where the probed
  values of $x_2\sim p_t/\sqrt{s}\,\exp(-y_h)$ are small but the
  predictions are clearly different. The slope of the cross section is
  much larger when described in our model as compared with the DHJ
  model, since $\gamma$ rises towards 1 much faster. Hence, a
  measurement of the slopes at moderate rapidities $y_h$ at LHC would
  allow a discrimination between the DHJ model and our model in a
  region where small-$x$ physics may be expected to be applicable.
  Since a logarithmic rise of $\gamma$ is a generic signature of BFKL
  evolution, these measurements offer the possibility of testing
  whether such small-$x$ evolution is actually relevant at present-day
  hadron colliders.

\begin{figure}[htb]
\centering
\includegraphics*[width=110mm]{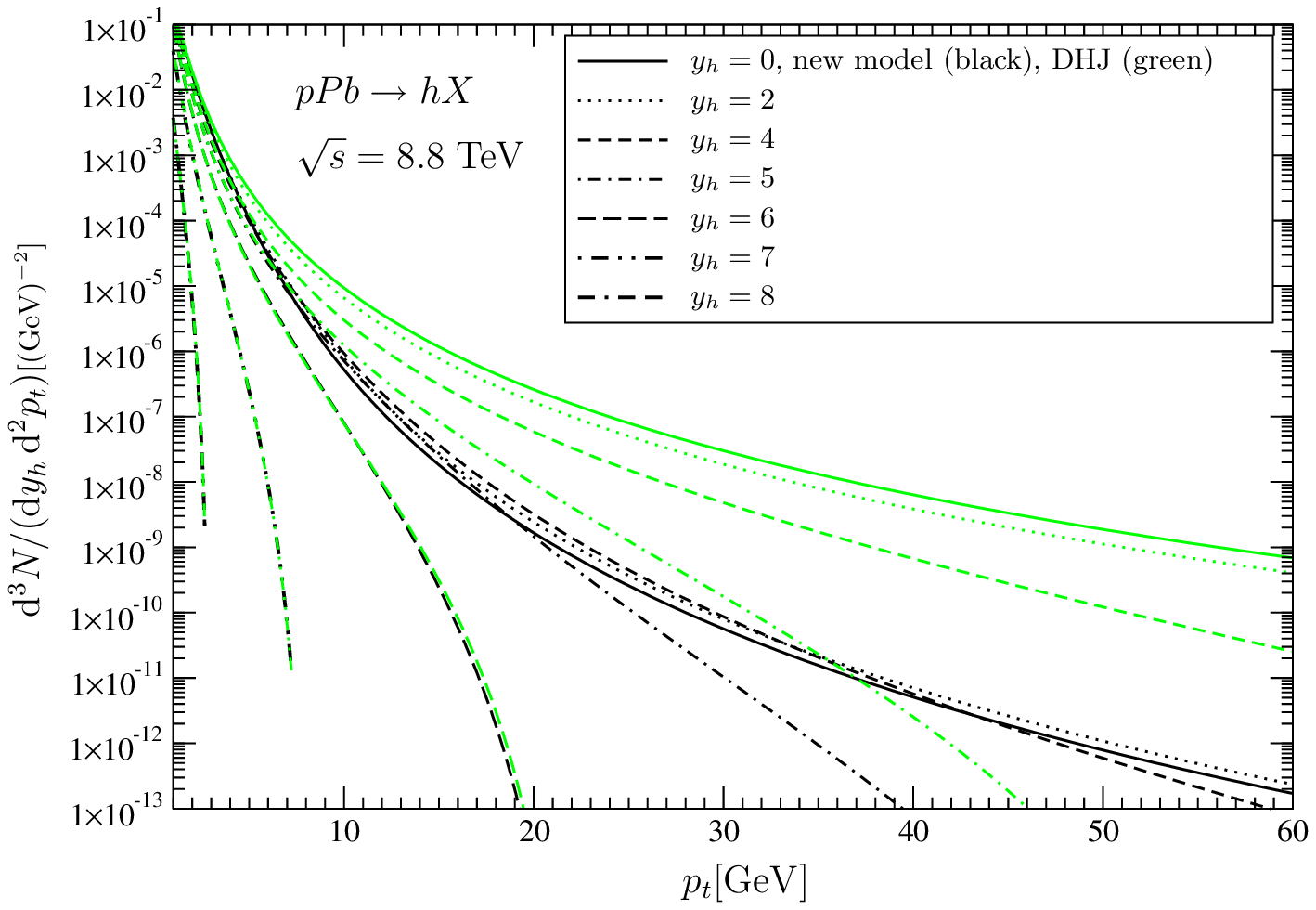}
\caption{\small\label{fig_pPb_LHC} Same as Fig.~\ref{fig_LHC}
  but for $p$-$Pb$ collisions at the LHC energy of $\sqrt{s}=8.8\;{\rm TeV}$. 
We have used $A_{\rm eff}=20$.}
\end{figure}

The $p$-$Pb$ predictions for LHC are very similar. However, due to
the smaller energy of $\sqrt{s}=8.8\;{\rm TeV}$ the predictions are
already comparable for smaller rapidities, i.e.\ for $y_h \approx 6$, cf.\ 
Fig.~\ref{fig_pPb_LHC}.
Here the rapidities are given for the nucleon-nucleon center of mass
frame, which for LHC is not the lab frame in contrast to RHIC. 
This means that in
terms of rapidities in the lab frame there is a slight offset of $\Delta y_h
= y_{\rm lab} - y_{cm} \approx 0.47$ to take into account.  

Note that for the whole kinematic range depicted in Figs.\ \ref{fig_LHC} and \ref{fig_pPb_LHC} 
the $x_2$ values are well below $0.01$. 

\subsection{Jet production}

Unlike in the case of the DIS cross section, geometric scaling of the
dipole amplitude does not lead to scaling of the hadron production
cross section at RHIC or LHC, because of the convolution of the
amplitude with the non-scaling parton distributions and fragmentation
functions. This effect can be reduced by considering jet production. The
description of the jet cross section does not involve any
fragmentation functions, but reduces to just a sum over products
of dipole amplitudes and parton distribution functions,
\begin{equation}
{dN_h \over dy_h d^2p_t}= 
{K(y_h) \over (2\pi)^2} \left[\sum_q f_{q/p}(x_F,p_t^2)\, N_F 
\left(p_t,x_2\right)+~
f_{g/p}(x_F,p_t^2)\, N_A \left(p_t,x_2\right)\right],
\end{equation}
where $x_{F}= p_t/\sqrt{s}\,\exp(y_h)$ and $x_2=x_F\exp(-2y_h)=
p_t/\sqrt{s}\,\exp(-y_h)$.  This means that in the kinematical regions
where either the gluon contribution or the quark contribution is
dominant (in general in a small kinematical region), the corresponding
distribution function can be divided out, so that one obtains the
dipole amplitude directly from the data. Of course, if the dipole
amplitude is only mildly scaling violating, this kinematical region
may be too small to observe the violations in this way. At LHC, the
gluon contribution to the jet cross section is reasonably dominant for
transverse momenta $p_t\lesssim 15$ GeV and hadron rapidities
$y_h=0-2$. In this region, the scaling violations of the ratio
  $(p_t^2\,dN_h/dy_h d^2p_t)/f_{g/p}(x_F,p_t^2)$ are in the DHJ model
  about 30\%, while the violations for the exactly scaling model
  (\ref{gamma_alpha}) are, due to quark contributions, still about
  10\%. We conclude that it may be difficult to attribute any
observed violations directly to $N_A$.  A similar conclusion holds for
$N_F$ in the region where quarks dominate (when $x_F \simorder 0.1$,
see Fig.\ \ref{fig_regions}) .

In summary, even for jet production, where there are no
complications from the fragmentation functions, it may not be possible to
establish geometric scaling violations conclusively due to the mixture of 
quark and gluon contributions. The kinematic range at LHC where either
quark or gluons dominate is probably too small to reach a definite 
conclusion about scaling violations.

\section{Conclusions}

We have presented a new phenomenological model of the dipole
scattering amplitude to demonstrate that the RHIC data for hadron
production in $d$-$Au$ collisions for all available rapidities are
compatible with geometric scaling. Moreover, the model also provides a
reasonable description of the small-$x$ DIS data.  On the other hand,
in a region of $y_h$ and $p_t$ for which the probed values of $x$ are
sufficiently small, the RHIC data are also compatible with geometric
scaling violating models, such as the DHJ model.  The fact that the
DHJ model, which incorporates scaling violations from BFKL (or more
generally BK) evolution to some extent, also describes the forward
RHIC data suggests that the data simply do not span a sufficiently
large region in $p_t$ and $y_h$ to demonstrate possible violations of
geometric scaling.  Hence, it cannot be concluded that scaling
violations of the dipole scattering amplitude play a role at RHIC.

The breakdown of the DHJ model at midrapidity might simply be due to
the probed values of $x$ being not sufficiently small.  The situation
is different at LHC in $p$-$p$ and $p$-$Pb$ collisions.  For smaller
rapidities, but still within the region where the small-$x$
description could be applicable, the DHJ model and the new scaling
model lead to different predictions for the $p_t$ fall-off of the
cross section.  This fall-off is determined by how fast the anomalous
dimension $\gamma$ approaches 1 for large transverse momentum. BFKL
evolution typically leads to a logarithmic rise of $\gamma \to 1$ with
transverse momentum and therefore implies a fall-off that is much
slower than one finds for the new scaling model that is compatible
with both the RHIC and the DIS data.  Therefore, at LHC in both
$p$-$p$ and $p$-$Pb$ collisions the transverse momentum distribution
will probe for the first time at sufficiently small $x$ the rise of
the anomalous dimension $\gamma$, and will thereby provide an
important test of the expectations from small-$x$ evolution.

\begin{acknowledgments}
  We thank Adrian Dumitru for helpful comments on the manuscript and
  Jamal Jalilian-Marian, Eric Laenen, Raimond Snellings and Werner
  Vogelsang for useful discussions. This research is part of the
  research program of the ``Stichting voor Fundamenteel Onderzoek der
  Materie (FOM)'', which is financially supported by the ``Nederlandse
  Organisatie voor Wetenschappelijk Onderzoek (NWO)''.
\end{acknowledgments}


\begin{thebibliography}{99}

\bibitem{Gribov:1984tu}
  L.~V.~Gribov, E.~M.~Levin and M.~G.~Ryskin,
  Phys.\ Rept.\  {\bf 100}, 1 (1983).

\bibitem{Laenen:1995fh}
  E.~Laenen and E.~Levin,
  Nucl.\ Phys.\ B {\bf 451}, 207 (1995).

\bibitem{Balitsky:1995ub}
  I.~Balitsky,
  Nucl.\ Phys.\ B {\bf 463}, 99 (1996).

\bibitem{Kovchegov:1999yj}
  Y.~V.~Kovchegov,
  Phys.\ Rev.\ D {\bf 60}, 034008 (1999).

\bibitem{Mueller:1989st}
  A.~H.~Mueller,
  Nucl.\ Phys.\ B {\bf 335}, 115 (1990).

\bibitem{Dumitru:2002qt}
  A.~Dumitru and J.~Jalilian-Marian,
  Phys.\ Rev.\ Lett.\  {\bf 89}, 022301 (2002).

\bibitem{Kharzeev:2002pc}
  D.~Kharzeev, E.~Levin and L.~McLerran,
  Phys.\ Lett.\  B {\bf 561}, 93 (2003).

\bibitem{Kharzeev:2003wz}
  D.~Kharzeev, Y.~V.~Kovchegov and K.~Tuchin,
  Phys.\ Rev.\  D {\bf 68}, 094013 (2003).

\bibitem{Albacete:2003iq}
  J.~L.~Albacete, N.~Armesto, A.~Kovner, C.~A.~Salgado and U.~A.~Wiedemann,
  Phys.\ Rev.\ Lett.\  {\bf 92}, 082001 (2004).

\bibitem{Baier:2003hr}
  R.~Baier, A.~Kovner and U.~A.~Wiedemann,
  Phys.\ Rev.\  D {\bf 68}, 054009 (2003).

\bibitem{Kharzeev:2002ei}
  D.~Kharzeev, E.~Levin and M.~Nardi,
  Nucl.\ Phys.\  A {\bf 730}, 448 (2004)
  [Erratum-ibid.\  A {\bf 743}, 329 (2004)].

\bibitem{JK}
  J.~Jalilian-Marian and Y.~V.~Kovchegov,
  Prog.\ Part.\ Nucl.\ Phys.\  {\bf 56}, 104 (2006).

\bibitem{GBW}
  K.~Golec-Biernat and M.~W\"usthoff,
  Phys.\ Rev.\ D {\bf 59}, 014017 (1999).

\bibitem{Stasto:2000er}
  A.~M.~Sta\'sto, K.~Golec-Biernat and J.~Kwieci\'nski,
  Phys.\ Rev.\ Lett.\  {\bf 86}, 596 (2001).

\bibitem{Gelis:2006bs}
  F.~Gelis, R.~Peschanski, G.~Soyez and L.~Schoeffel,
  Phys.\ Lett.\  B {\bf 647}, 376 (2007).

\bibitem{Bartels:2002cj}
  J.~Bartels, K.~Golec-Biernat and H.~Kowalski,
  Phys.\ Rev.\ D {\bf 66}, 014001 (2002).

\bibitem{Kwiecinski:2002ep}
  J.~Kwieci\'nski and A.~M.~Sta\'sto,
  Phys.\ Rev.\  D {\bf 66}, 014013 (2002).

\bibitem{Kuraev:1977fs}
  E.~A.~Kuraev, L.~N.~Lipatov and V.~S.~Fadin,
  Sov.\ Phys.\ JETP {\bf 45}, 199 (1977)
  [Zh.\ Eksp.\ Teor.\ Fiz.\  {\bf 72}, 377 (1977)].
  
\bibitem{Balitsky:1978ic}
  I.~I.~Balitsky and L.~N.~Lipatov,
  Sov.\ J.\ Nucl.\ Phys.\  {\bf 28}, 822 (1978)
  [Yad.\ Fiz.\  {\bf 28}, 1597 (1978)].

\bibitem{LevinTuchin}
  E.~Levin and K.~Tuchin,
  Nucl.\ Phys.\ A {\bf 691}, 779 (2001).

\bibitem{MuellerTr}
  A.~H.~Mueller and D.~N.~Triantafyllopoulos,
  Nucl.\ Phys.\ B {\bf 640}, 331 (2002).

\bibitem{Triantafyllopoulos:2002nz}
  D.~N.~Triantafyllopoulos,
  Nucl.\ Phys.\ B {\bf 648}, 293 (2003).

\bibitem{IIM2}
  E.~Iancu, K.~Itakura and L.~McLerran,
  Nucl.\ Phys.\ A {\bf 708}, 327 (2002).

\bibitem{IIM}
  E.~Iancu, K.~Itakura and S.~Munier,
  Phys.\ Lett.\ B {\bf 590}, 199 (2004).

\bibitem{Gotsman:2002yy}
  E.~Gotsman, E.~Levin, M.~Lublinsky and U.~Maor,
  Eur.\ Phys.\ J.\  C {\bf 27}, 411 (2003).

\bibitem{DHJ1}
  A.~Dumitru, A.~Hayashigaki and J.~Jalilian-Marian,
  Nucl.\ Phys.\ A {\bf 765}, 464 (2006).

\bibitem{DHJ2}
  A.~Dumitru, A.~Hayashigaki and J.~Jalilian-Marian,
  Nucl.\ Phys.\ A {\bf 770}, 57 (2006).

\bibitem{KKT}
  D.~Kharzeev, Y.V.~Kovchegov and K.~Tuchin,
  Phys.\ Lett.\ B {\bf 599}, 23 (2004).

\bibitem{BDH}
  D.~Boer, A.~Dumitru and A.~Hayashigaki,
  Phys.\ Rev.\ D {\bf 74}, 074018 (2006).

\bibitem{Jager:2002xm}
  B.~J\"ager, A.~Sch\"afer, M.~Stratmann and W.~Vogelsang,
  Phys.\ Rev.\  D {\bf 67}, 054005 (2003).

\bibitem{BUW}
  D.~Boer, A.~Utermann and E.~Wessels,
  Phys.\ Rev.\  D {\bf 75}, 094022 (2007). 

\bibitem{Iancu:2006uc}
  E.~Iancu, C.~Marquet and G.~Soyez,
  Nucl.\ Phys.\  A {\bf 780}, 52 (2006).

\bibitem{MV}
  L.~McLerran and R.~Venugopalan, Phys.\ Rev.\ D {\bf 49}, 2233 (1994);
  ibid.\ {\bf 49}, 3352 (1994);
  Y.~V.~Kovchegov, ibid.\ {\bf 54}, 5463 (1996);
  ibid.\ {\bf 55}, 5445 (1997).

\bibitem{Adams:2003im}
  J.~Adams {\it et al.}  [STAR Collaboration],
  Phys.\ Rev.\ Lett.\  {\bf 91}, 072304 (2003).

\bibitem{Arsene:2004ux}
  I.~Arsene {\it et al.}  [BRAHMS Collaboration],
  Phys.\ Rev.\ Lett.\  {\bf 93}, 242303 (2004).

\bibitem{Adams:2006uz}
  J.~Adams {\it et al.}  [STAR Collaboration],
  Phys.\ Rev.\ Lett.\  {\bf 97}, 152302 (2006).

\bibitem{Lai:1999wy}
  H.~L.~Lai {\it et al.}  [CTEQ Collaboration],
  Eur.\ Phys.\ J.\  C {\bf 12}, 375 (2000).

\bibitem{Kniehl:2000fe}
  B.~A.~Kniehl, G.~Kramer and B.~P\"otter,
  Nucl.\ Phys.\  B {\bf 582}, 514 (2000).

\bibitem{Adloff:2000qk}
  C.~Adloff {\it et al.}  [H1 Collaboration],
  Eur.\ Phys.\ J.\  C {\bf 21}, 33 (2001).

\bibitem{Breitweg:2000yn}
  J.~Breitweg {\it et al.}  [ZEUS Collaboration],
  Phys.\ Lett.\  B {\bf 487}, 53 (2000).

\bibitem{Chekanov:2001qu}
  S.~Chekanov {\it et al.}  [ZEUS Collaboration],
  Eur.\ Phys.\ J.\  C {\bf 21}, 443 (2001).

\bibitem{Avsar:2007ht}
  E.~Avsar and G.~Gustafson,
  JHEP {\bf 0704}, 067 (2007).

\end{thebibliography}
\end{document}